\documentclass[conference]{IEEEtran}
\usepackage{graphicx}%
\usepackage{bm}%
\usepackage{amsfonts}
\usepackage{amssymb}
\usepackage{times}
\usepackage{subfigure}
\usepackage{enumitem}
\usepackage{amsmath} %Latex
\usepackage{latexsym}
\usepackage{hhline}
\usepackage{cite}
\usepackage{caption}
\usepackage{algorithm}
\usepackage{algpseudocode}
\algrenewcommand\algorithmicrequire{\textbf{Input:}}
\algrenewcommand\algorithmicensure{\textbf{Output:}}
\usepackage{multirow} %multirow for format of table
\usepackage{xcolor}
\usepackage{epstopdf}
\usepackage{aligned-overset}

\allowdisplaybreaks

\usepackage{geometry}
\begin{document}
\bstctlcite{bibfbcs:BSTcontrol}
\title{\vspace{-2.2mm} A Novel Cross-Domain Channel Estimation Scheme for OFDM \vspace{-2.9mm}}
\author{\IEEEauthorblockN{
Mingcheng Nie\IEEEauthorrefmark{1},
Ruoxi Chong\IEEEauthorrefmark{2}, 
Shuangyang Li\IEEEauthorrefmark{3}, 
Weijie Yuan\IEEEauthorrefmark{4}, \\
Derrick Wing Kwan Ng\IEEEauthorrefmark{5},
Michail Matthaiou\IEEEauthorrefmark{2}, 
Giuseppe Caire\IEEEauthorrefmark{3}, 
and
Yonghui Li\IEEEauthorrefmark{1}
\vspace{2mm}
}
\IEEEauthorblockA{
\IEEEauthorrefmark{1}The University of Sydney, Sydney, Australia\\
\IEEEauthorrefmark{2}Queen’s University Belfast, Belfast, U.K.\\
\IEEEauthorrefmark{3}Technische Universit{\"a}t Berlin, Berlin, Germany\\
\IEEEauthorrefmark{4}Southern University of Science and Technology, Shenzhen, China\\
\IEEEauthorrefmark{5}University of New South Wales, Sydney, Australia \vspace{-7mm}
}}

%Emails: mingcheng.nie@sydney.edu.au, shuangyang.li@uwa.edu.au, caire@tu-berlin.de, yonghui.li@sydney.edu.au

\maketitle
\begin{abstract}
In this paper, we propose a novel cross-domain channel estimation (CDCE) algorithm for orthogonal frequency division multiplexing (OFDM) systems, leveraging the unique characteristics of the delay-Doppler (DD) domain channel. Specifically, the proposed algorithm 
transforms the time-frequency (TF) domain pilot sequence of OFDM into the DD domain and applies a two-dimensional (2D) twisted-convolution for acquiring a coarse estimation of the underlying channel delay and Doppler. Then, the OFDM channel estimation is formulated as a sparse signal recovery problem in the TF domain according to the dictionary derived based on the obtained delay and Doppler estimates. Furthermore, a low-complexity $\ell_1$-regularized least-square estimator is proposed to effectively solve this problem.
Moreover, we further develop a performance analysis framework of the proposed scheme based on the ambiguity function (AF) of the adopted pilot sequence. 
Our numerical results demonstrate noticeable estimation performance improvement compared to conventional OFDM channel estimation methods, particularly in the presence of high channel mobility.

\let\thefootnote\relax\footnotetext{
The work of M. Matthaiou was supported by a research grant from the European Research Council (ERC) under the European Union’s Horizon 2020 research and innovation programme (grant No. 101001331).   }
\end{abstract}

\vspace{-2mm}
\section{Introduction}
Future wireless networks are expected to provide robust communication services in challenging scenarios, including high-mobility environments. In such cases, the Doppler effect caused by the channel is often non-negligible, posing a significant challenge to the currently deployed orthogonal frequency division multiplexing (OFDM) modulation scheme~\cite{yuan2023new}. Indeed, the success of OFDM relies heavily on the assumption of perfect subcarrier orthogonality, which becomes overly idealistic in transmissions with high channel mobility~\cite{li2021performance,nie2024uplink}. In such cases, the large Doppler spread introduces non-uniform frequency shifts across subcarriers, thereby undermining subcarrier orthogonality. Consequently, conventional OFDM systems suffer from severe inter-carrier interference (ICI) and reduced channel coherence time, thereby significantly degrading the accuracy of channel estimation.
Meanwhile, the drive to maintain high spectral efficiency limits the number of pilots that can be allocated for channel estimation, causing conventional time-frequency (TF) domain estimation methods to suffer notable performance degradation under such demanding conditions.

Recently, delay-Doppler (DD) waveforms, such as orthogonal time-frequency space (OTFS)~\cite{hadani2017orthogonal}, have attracted growing attention due to their remarkable capability to provide reliable communication over high-mobility channels, which is achieved by exploiting the appealing DD domain channel characteristics. 
Specifically, the DD domain representation of a time-varying channel is quasi-static, compact, and potentially sparse, which can underpin novel low-complexity
channel estimation schemes. For example, a typical DD domain channel estimation scheme considers an embedded pilot in the DD domain with a sufficiently large guard space around it~\cite{raviteja2019embedded}, where the DD domain channel can be estimated by comparing the received pilot symbols with a given threshold. However, this estimation requires a DD domain pilot that spreads across the entire TF domain, which is practically unachievable in conventional OFDM systems.
This naturally raises a practical question: Can we still leverage the benefits of DD domain channels while employing OFDM pilots transmitted in the TF domain? Providing an effective solution to this question could dramatically enhance the channel estimation performance in conventional OFDM systems without the need to transmit DD domain waveforms, such as OTFS, thereby avoiding additional hardware complexity and cost.

Leveraging the insights from the DD domain, previous works have focused on enhancing OTFS performance. For instance, in~\cite{yuan2021data}, a superimposed pilot scheme was proposed to enhance the spectral efficiency by embedding pilots directly into data symbols. The resulting pilot-data interference was managed via an iterative method that combines threshold-based channel estimation, interference cancellation, and data detection, while adaptively updating the threshold based on symbol reliability. Furthermore, the authors in~\cite{gaudio2021otfs} introduced a two-stage approach estimation, which first coarsely estimates the on-grid delay and Doppler shifts via thresholding, and a low-complexity maximum likelihood estimator was used to refine the estimates. Moreover, \cite{he2024pilot} proposed a novel DD domain pilot sequence design by minimizing the Cram$\Acute{\text{e}}$r-Rao bound (CRB), explicitly accounting for windowing effects and data interference. However, a comprehensive solution that directly integrates the true structure of OFDM with cross-domain processing has not yet been fully realized. Specifically, to the best of our knowledge, there has been no approach that simultaneously considers the TF and DD domains in a unified framework, while maintaining the core OFDM structure.

In this paper, we aim to bridge this gap by proposing a novel cross-domain channel estimation (CDCE) method that builds upon the OFDM framework by integrating both the TF domain and the DD domain. Our method leverages the major advantages of DD domain twisted-convolution to distinguish and extract channel delay and Doppler shifts, which are fed back to the TF domain to formulate an $\ell_1$-regularized least-square minimization problem for fading coefficients' estimation. The overall approach has low-complexity since it does not require any matrix inversions or exhaustive searches, compared to conventional methods. This allows us to exploit the unique properties of the DD domain for better management of the Doppler effects while retaining the structure of the TF domain for efficient data transmission. Through simulations, we demonstrate that the proposed method significantly improves the channel estimation accuracy and achieves a lower normalized mean square error (NMSE) compared to conventional OFDM estimation schemes.

\emph{Notation:}
The superscripts $(\cdot)^{\rm{H}}$, $(\cdot)^{*}$ and $(\cdot)^{\rm{T}}$ denote the Hermitian transpose, conjugate and transpose operation, respectively; 
${\rm Tr}\{\cdot\}$ and ${\rm{vec}}\left( \cdot \right)$ denote the trace and the vectorization operation of a matrix; 
%${{\mathop{\rm cov}} \left( \bf x\right)}$ denotes the covariance matrix of $\bf x$;
%$f^{*}(\cdot)$ denotes the conjugate of $f(\cdot)$; 
${{\bf{F}}_N}$ denotes the normalized discrete Fourier transform (DFT) matrix of size $N\times N$;
${\bf I}_M$ represents the $M\times M$ identity matrix;
``$ \otimes $" denotes the Kronecker product operator. 
$\mathbb{E}[\cdot]$ denotes the expectation operation;
$\mathbf{X}=\mathrm{diag}\left(\mathbf{x}\right)$ denotes a diagnol matrix $\mathbf{X}$ whose diagonal entries are given by vector $\mathbf{x}$.
%$|\cdot|$ returns the cardinality of a set;
%$\textrm{det}(\cdot)$ and $\rm{Tr}(\cdot)$ denote the determinant and trace operations of a matrix, respectively; 
%$\simeq$ denotes asymptotically equal;
%$\mathbb{E}[\cdot]$ denotes the statistical expectation. 
The circular symmetric complex Gaussian distribution having variance $\sigma^2$ is denoted by $\mathcal{CN}(0,\sigma^2)$. Finally, ${\mathbb{C}}$  denotes the complex number field.

\vspace{-2mm}

\section{System Model}
\vspace{-1mm}
% \subsection{OFDM Transmissions over Doubly-Selective Channel}
We consider an OFDM system with $M$ subcarriers and $N$ time slots, where critical sampling is assumed~\cite{li2020code} such that the subcarrier spacing $\Delta f$ and the time slot duration $T$ satisfy $\Delta f \cdot T =1$.
Let $\mathbf{X}_\mathrm{TF}$ be the TF domain transmitted symbol matrix of dimension $M \times N$. The corresponding time domain transmitted symbol matrix $\mathbf{S}\in\mathbb{C}^{M\times N}$ can then be obtained by applying the inverse discrete Fourier transform (IDFT) along the frequency dimension as follows:
\begin{align}
    \mathbf{S}=\mathbf{F}_M^{\rm H}\mathbf{X}_\mathrm{TF}. \label{time domain S}
\end{align}
By appending the OFDM cyclic prefix (CP) to $\mathbf{S}$, we have
\begin{align}
    \tilde{\mathbf{S}}=\mathbf{A}_{\mathrm{CP}}\mathbf{S}=\mathbf{A}_{\mathrm{CP}}\mathbf{F}_M^{\rm H}\mathbf{X}_\mathrm{TF},
\end{align}
which can also be written in the vector form as
\begin{align}
    \tilde{\mathbf{s}}=(\mathbf{I}_N\otimes\mathbf{A}_{\mathrm{CP}})\mathbf{s}=(\mathbf{I}_N\otimes\mathbf{A}_{\mathrm{CP}}\mathbf{F}_M^{\rm H})\mathbf{x}_\mathrm{TF}.
\end{align}
Here, $\mathbf{A}_{\mathrm{CP}}=[\mathbf{G}_{\mathrm{CP}},\mathbf{I}_{M}]^T\in\mathbb{R}^{(M+L_{\mathrm{CP}})\times M}$ is the CP addition matrix, where $\mathbf{G}_{\mathrm{CP}}$ of size $M\times L_{\mathrm{CP}}$ includes the last $L_{\mathrm{CP}}$ columns of the identity matrix $\mathbf{I}_{M}$, and $L_{\mathrm{CP}}$ denotes the length of the CP, whose value is determined by the maximum delay of the channel. 
Note that after inserting the CP, we have $\tilde{\mathbf{s}}\in\mathbb{C}^{(M+L_{\rm CP})N\times 1}$.
The continuous time domain transmit signal is obtained by applying the transmitter pulse shaping $p(t)$ to $\tilde{\mathbf{s}}$, given by \vspace{-0.7mm}
\begin{align}
    s(t)=\sum\nolimits_{n=-NL_{\mathrm{CP}}}^{MN}\tilde{{s}}[n] p(t-nT_s),
\end{align}
where $\tilde{{s}}[n]$ denotes the $n$-th element of $\tilde{\mathbf{s}}$, and $T_s=\frac{T}{M}$ is the time domain sampling period, which also aligns with the delay resolution.
Without loss of generality, we consider a time-varying channel, whose DD domain representation can be written as
\begin{align}
    h(\tau,\nu)=\sum\nolimits_{p=1}^{P}h_p\delta(\tau-\tau_p)\delta(\nu-\nu_p),
\end{align}
where $h_{p} \in \mathbb C$, $\tau_{p}=\left(l_{p}+\imath_{p}\right)\frac{1}{M\Delta f}$, and $\nu_{p}=\left(k_{p}+\kappa_{p}\right)\frac{1}{NT}$ denote the channel coefficient, delay, and Doppler shifts associated with $p$-th path, respectively. Here, $l_{p}$ and $k_{p}$ represent the integer delay and integer Doppler indices, respectively, while $-{1}/{2}\le\imath_p\le {1}/{2}$ and $-{1}/{2}\le\kappa_p\le {1}/{2}$ represent the fractional delay and fractional Doppler indices, respectively. %
%In this paper, we consider the wireless channel with sufficient delay and Doppler resolutions, i.e., the fractional delay and Doppler indices are of zero values. 
After passing through the time-varying channel, the continuous time domain received signal $r(t)$ can be expressed by
\begin{align}
    r(t)=\sum\nolimits_{p=1}^{P} h_p s(t-\tau_p)e^{j2\pi\nu_p(t-\tau_p)}+w(t),
\end{align}
where $w(t)$ is the additive white Gaussian noise (AWGN) with the one-sided power spectral density $N_0$. By adopting a matched filter for the received signal, we obtain its discrete-time representation $r[m],-L_{\mathrm{CP}}\le m\le MN$, whose $m$-th element can be expressed as
\begin{align}
     \tilde{r}[m] 
     &=  \int_{-\infty}^{\infty} r(t) p^*(t-mT_s) dt+w[m]\notag\\
    &=\sum\nolimits_{n=-NL_{\mathrm{CP}}}^{MN}\tilde{{s}}[n]  g[m,n]+w[m].\label{g_first}
\end{align}
Specifically, $g_{[m,n]}$ is the $\left( {m,n} \right)$-th element of the time domain channel matrix ${\bf{G}}\in\mathbb{C}^{(M+L_{\rm {CP}})N\times (M+L_{\rm {CP}})N}$, which can be represented by
\begin{align}
    g[m,n] =\sum\nolimits_{p=1}^{P} h_p e^{j2\pi n \nu_p T_s} A^*((n-m)T_s+\tau_p,\nu_p).\label{g_element}
\end{align}
Moreover, $A(\tau,\nu)$ in (\ref{g_element}) denotes the ambiguity function (AF) of pulse $p(t)$ with respect to delay $\tau$ and Doppler shift $\nu$, which is defined by
\begin{align}
    A(\tau_p,\nu_p)& \triangleq\int_{-\infty}^{\infty} p(t) p^*(t-\tau_p) e^{-j2\pi\nu_p(t-\tau_p)} dt.
\end{align}
By stacking all the symbols into a vector, we have 
\begin{align}
\tilde{\mathbf{r}}&=\sum\nolimits_{p=1}^{P}\mathbf{G}_p\tilde{\mathbf{s}}+\mathbf{w}
=\mathbf{G}\tilde{\mathbf{s}}+\mathbf{w},
\end{align}
where $\mathbf{G}_p$ denotes the $p$-th resolvable path component of the time domain channel ${\bf {G}}$.
We then remove the CP in the time domain by applying the CP reduction matrix $\mathbf{R}_{\mathrm{CP}} \triangleq [\mathbf{0}_{M\times L_{\mathrm{CP}}}, \mathbf{I}_{M}]$ to $\tilde{\mathbf{r}}$, resulting in the time-domain received signal $\mathbf{r}$ that is therefore given by
\begin{align}
    \mathbf{r}&=\left(\mathbf{I}_N\otimes\mathbf{R}_{\mathrm{CP}}\right)\mathbf{G}(\mathbf{I}_N\otimes\mathbf{A}_{\mathrm{CP}})\mathbf{s}+\mathbf{w}=\mathbf{G}_{\mathrm{T}}\mathbf{s}+\mathbf{w},\label{time_domain_received_noCP2}
\end{align}
where $\mathbf{w}$ denotes the additive noise vector and $\mathbf{G}_{\rm T}\in\mathbb{C}^{MN\times MN}$ represents the effective time domain channel matrix after removing the CP.
Note that in~\eqref{time_domain_received_noCP2} and what follows, we slightly abuse our notations for simplicity and adopt $\bf w$ for representing the noise vector.
The TF domain receive symbols are then obtained by applying an DFT to the time domain symbol vector $\mathbf{r}$, leading to the TF domain input-output relationship as
\begin{align}
    \mathbf{y}_\mathrm{TF}
    =&\left(\mathbf{I}_N\otimes\mathbf{F}_M\right) \mathbf{G}_{\mathrm{T}}  \left(\mathbf{I}_N\otimes\mathbf{F}_M^{\rm H}\right)\mathbf{x}_\mathrm{TF}+\mathbf{w}\notag \\
    =&\mathbf{H}_\mathrm{TF}\mathbf{x}_\mathrm{TF}+\mathbf{w},\label{IO_TF}
\end{align}
where $\mathbf{H}_\mathrm{TF}\in\mathbb{C}^{MN\times MN}$ represents the effective TF domain channel matrix. 

\vspace{-1.5mm}
\section{Cross-Domain Channel Estimation}
The proposed CDCE algorithm leverages the unique convolutional structure of the effective DD domain channel.
Specifically, building upon the system model in~\eqref{IO_TF}, we first perform a coarse estimation of the channel delay and Doppler coefficients in the DD domain, and then estimate the fading coefficients in the TF domain capitalizing on the previously obtained delay and Doppler estimates.

For ease of our discussion, we first present the equivalent input-output relationship in the DD domain, which can be obtained via the Symplectic Finite Fourier Transform (SFFT), denoted by~\cite{wei2021orthogonal} 
\begin{align}
    {\mathbf{y}}_\mathrm{DD}& = \left(\mathbf{F}_N\otimes\mathbf{F}_M^{\rm H}\right)\mathbf{y}_{\mathrm{TF}}\nonumber\\
    &=\left(\mathbf{F}_N\otimes\mathbf{F}_M^{\rm H}\right)\mathbf{H}_\mathrm{TF}\left(\mathbf{F}_N^{\rm H}\otimes\mathbf{F}_M\right)\mathbf{x}_{\mathrm{DD}}+\mathbf{w}\notag\\
    &= \left(\mathbf{F}_N\otimes\mathbf{R}_{\mathrm{CP}}\right) \mathbf{G}\left(\mathbf{F}_N^{\rm H}\otimes\mathbf{A}_{\mathrm{CP}}\right)\mathbf{x}_{\mathrm{DD}}+\mathbf{w}\notag\\
    &=\left(\mathbf{I}_N\otimes\mathbf{R}_{\mathrm{CP}}\right)\tilde{\mathbf{H}}_\mathrm{DD}\left(\mathbf{I}_N\otimes\mathbf{A}_{\mathrm{CP}}\right)\mathbf{x}_{\mathrm{DD}}+\mathbf{w}\label{IO_DD_data0}\\
    &={\mathbf{H}}_{\mathrm{DD}}\mathbf{x}_{\mathrm{DD}}+\mathbf{w}
    ,\label{IO_DD_data}
   % &=\tilde{\mathbf{H}}_{\mathrm{DD}}(\mathbf{x}^{\mathrm{pilot}}_{\mathrm{DD}}+\mathbf{x}^{\mathrm{data}}_{\mathrm{DD}})+\mathbf{w}.\label{IO_DD_data}
\end{align}
where ${\bf x}_{\rm DD} \triangleq \left(\mathbf{F}_N\otimes\mathbf{F}_M^{\rm H}\right)\mathbf{x}_{\mathrm{TF}}$
% \begin{align}
%     {\bf x}_{\rm DD} \triangleq \left(\mathbf{F}_N\otimes\mathbf{F}_M^{\rm H}\right)\mathbf{x}_{\mathrm{TF}},
% \end{align}
is the equivalent DD domain presentation  corresponding to $\mathbf{x}_{\mathrm{TF}}$. 
In~\eqref{IO_DD_data0} and~\eqref{IO_DD_data}, 
$\tilde{\mathbf{H}}_\mathrm{DD}\in\mathbb{C}^{(M+L_{\rm {CP}})N\times (M+L_{\rm {CP}})N}$ and ${\mathbf{H}}_{\mathrm{DD}}\in\mathbb{C}^{MN\times MN}$ are the equivalent DD domain channel matrix before and after CP removal, respectively.
It is important to note that the addition and removal of the CP constitute a unitary operation, which affects only the dimensions of the channel matrices by appending or discarding a limited number of rows or columns. Consequently, the underlying delay, Doppler, and fading coefficients of the channel responses remain unchanged before and after this process.

\vspace{-1mm}
\subsection{DD Domain Delay and Doppler Estimation}
\vspace{-1mm}
Based on the DD domain input-output relationship, we first estimate the delay and Doppler coefficients of each resolvable path in the DD domain.
In fact, directly estimating these parameters in the TF domain may be challenging due to the 
ICI and inter-symbol interference (ISI). 
Therefore, motivated by the pulse compression in~\cite{zhang2023radar}, we consider estimating the delay and Doppler coefficients by exploiting the twisted-correlation between the received signal $\mathbf{y}_\mathrm{DD}$ and the transmitted signal $\mathbf{x}_\mathrm{DD}$.

To do so, we first define the estimated delay and Doppler indices vectors as $\hat{\boldsymbol{l}}=[\hat{l}_1,\dots,\hat{l}_{\hat{P}}]$ and $\hat{\boldsymbol{k}}=[\hat{k}_1,\dots,\hat{k}_{\hat{P}}]$, respectively,
where $\hat{P}$ denotes the estimated number of channel paths that is usually chosen sufficiently large such that $\hat P \ge P$. 
Let the matrix $\mathbf{V}_{\mathrm{DD}}$ be the \textit{twisted-convolution matrix}, whose $(l,k)$-th entry represents the likelihood of the underlying delay and Doppler pair. Specifically, we have
\begin{align}
 V_{\mathrm{DD}}[l,k] &= 
\sum_{m=0}^{M-1}\;
\sum_{n=0}^{N-1}
Y_{\mathrm{DD}}^{*}[m,n]\,
X_{\mathrm{DD}}\!\bigl[[m-l]_{M},\, [n-k]_{N}\bigr]\nonumber\\
& \qquad\qquad \qquad \ \times \alpha[m-l,n-k]  e^{j2\pi\frac{k(m-l)}{MN}},\label{DD_cross_corr}
\end{align}
where $l \in [0, M-1]$, $k \in [0, N-1]$, and $[\cdot]_M$ and $[\cdot]_N$ denote the modulo-$M$ and modulo-$N$ operations, respectively. The matrices $\mathbf{Y}_\mathrm{DD}\in\mathbb{C}^{M\times N}$ and $\mathbf{X}_\mathrm{DD}\in\mathbb{C}^{M\times N}$ are reshaped versions of the vectors $\mathbf{y}_\mathrm{DD}$ and $\mathbf{x}_\mathrm{DD}$, with $Y_{\mathrm{DD}}[m,n]$ and $X_{\mathrm{DD}}[m,n]$ representing their $(m,n)$-th entry. The term $\alpha[l'-l,k'-k]$ is the phase factor introduced by twisted-convolution nature of DD domain communications, whose expression is given by
\begin{align}
    \alpha[m-l,n-k] =
    \begin{cases}
        e^{-j2\pi\frac{n-k}{N}}, &l'-l<0,\\
        1, &l'-l\ge 0.
    \end{cases}\label{auto_corr}
\end{align}

After computing the twisted-convolution matrix $\mathbf{V}_{\mathrm{DD}}$, a thresholding step is applied: any entry $V_{\mathrm{DD}}[l,k]$ with magnitude less than a threshold $\gamma$ is set to zero. In addition, all entries outside a target region $\mathcal{R} = [0, l_{\max}] \times [-k_{\max}, k_{\max}]$ are also zeroed out, where $l_{\max}$ and $k_{\max}$ denote the maximum delay and Doppler shifts of the channel typically known before the transmission. Finally, the delay and Doppler indices corresponding to the nonzero entries in $\mathbf{V}_{\mathrm{DD}}$ are collected, forming the vectors $\hat{\boldsymbol{l}}$ and $\hat{\boldsymbol{k}}$, respectively.

\vspace{-1mm}
\subsection{TF Domain Channel Coefficient Estimation}
\vspace{-1mm}
In the next step, we estimate the channel fading coefficients with consideration of the previously obtained delay and Doppler estimates from the DD domain.
Particularly, let us rewrite the TF domain input-output relationship in~\eqref{IO_TF} by
%The TF domain input-output relationship in (\ref{IO_TF}) can be rewritten by
\begin{align}
    \mathbf{y}_\mathrm{TF}=&\left(\mathbf{I}_N\otimes\mathbf{F}_M\right) \mathbf{G}_{\mathrm{T}}  \left(\mathbf{I}_N\otimes\mathbf{F}_M^{\rm H}\right)\mathbf{x}_\mathrm{TF}+\mathbf{w}\notag \\
    =&\left(\mathbf{I}_N\otimes\left(\mathbf{F}_M\mathbf{R}_{\mathrm{CP}}\right)\right) \mathbf{G} \left(\mathbf{I}_N\otimes\left(\mathbf{A}_{\mathrm{CP}}\mathbf{F}_M^{\rm H}\right)\right)\mathbf{x}_\mathrm{TF}+\mathbf{w}\notag\\
    =&\left(\mathbf{x}^{\rm T}_\mathrm{TF}\otimes\mathbf{I}_{MN}\right)\nonumber\\
    &\left( \left(\mathbf{I}_N\otimes\left(\mathbf{A}_{\mathrm{CP}}\mathbf{F}_M^{\rm H}\right)\right)^{\rm T}\otimes \left(\mathbf{I}_N\otimes\left(\mathbf{F}_M\mathbf{R}_{\mathrm{CP}}\right)\right) \right)\mathrm{vec}\left(\mathbf{G}\right)\notag\\
    =&\tilde{\mathbf{X}}_{\mathrm{TF}}\mathrm{vec}\left(\mathbf{G}\right)+\mathbf{w}=\tilde{\mathbf{X}}_{\mathrm{TF}}\boldsymbol{\Phi}(\boldsymbol{\tau},\boldsymbol{\nu})\boldsymbol{h}+\mathbf{w}\notag\\
    =&\mathbf{D}\boldsymbol{h}+\mathbf{w}.\label{IO_TF_Xh}
\end{align}
Here, we use $\boldsymbol{h}\in\mathbb{C}^{{\hat P}\times 1}$ to denote the fading vector of all paths and the $(i,j)$-th element of $\boldsymbol{\Phi}(\boldsymbol{\tau},\boldsymbol{\nu})\in\mathbb{C}^{(M+L_{\mathrm{CP}})^2N^2\times {\hat P}}$ is given by
\begin{align}
    \phi[i,j] = e^{j2\pi n' \nu_p T_s} A^*((n'-m')T_s+\tau_j,\nu_j),
\end{align}
where $m'=\left[i\right]_{(M+L_{\mathrm{CP}})N}$ and $n'=\lfloor\frac{i}{(M+L_{\mathrm{CP}})N}\rfloor$. 
With the DD domain estimated delay and Doppler shifts, i.e., $\hat{\boldsymbol{l}}$ and $\hat{\boldsymbol{k}}$, and the input-output relationship in (\ref{IO_TF_Xh}), the fading coefficients can be estimated in the TF domain by solving a least squares (LS) problem. Notice that the choice of the LS estimator depends on the dimension ratio of the dictionary matrix $\mathbf{D}$. In other words, the number of observations utilized and the dimension of the DD domain estimation output jointly determine the appropriate estimator. In particular, when $\mathbf{D}$ is a well-conditioned tall matrix, the LS problem can be solved directly using a pseudo-inverse of $\mathbf{D}$, i.e.,
\begin{align}
    \hat{\boldsymbol{h}}=\left(\mathbf{D}^{\mathrm H}\mathbf{D}\right)^{-1}\mathbf{D}^{\mathrm H}\mathbf{y}_\mathrm{TF}.
\end{align}
On the other hand, when $\mathbf{D}$ is a fat matrix, the LS problem can be solved via a least absolute shrinkage and selection operator (LASSO) optimization:
\begin{align}
    \hat{\boldsymbol{h}}=\arg\min_{\boldsymbol{h}} ||\mathbf{y}_{\mathrm{TF}}-\mathbf{D}\boldsymbol{h}||^2_2+\lambda ||\boldsymbol{h}||_1, \label{lasso_def}
\end{align}
where $||\cdot||_{b}$ denotes the norm-$b$ operation and $\lambda$ is the regularization parameter for the LASSO problem. It is worth noticing that LASSO applies an $\ell_1$-norm regularization, which enhances the sparsity  
 in the estimated TF domain fading vector $\hat{\boldsymbol{h}}$. An important observation is that the initial DD domain correlation output may include false estimates of the delay and Doppler shifts, while the subsequent LASSO-based estimation in the TF domain helps suppress these false estimates by driving the corresponding entries in $\hat{\boldsymbol{h}}$ to zero. In other words, all the entries set to zero by LASSO effectively indicate the false estimates from the DD domain. 
 
 In the following, we detail our procedure for solving the LASSO problem in (\ref{lasso_def}). Prior to the main iterations, we perform the following initializations:
\begin{subequations}\label{O1}
\begin{align}
    \epsilon&=1/||\mathbf{D}||_2^2, \ \gamma_{\mathrm{LASSO}}=\lambda \epsilon, \ \beta^0=1,\\
    \hat{\boldsymbol{h}}^0&=[0,\dots,0]^T, \ \mathbf{z}^0=\hat{\boldsymbol{h}}, 
\end{align}\label{init_para}
\end{subequations}\vspace{-4mm}

\noindent where $\epsilon$, $\gamma_{\mathrm{LASSO}}$, and $\beta$ denote the gradient-step size, threshold of LASSO, and Nesterov momentum parameter, respectively. Note that $\hat{\boldsymbol{h}}$ is initialized as a zero vector of length $\hat{P}$ while $\mathbf{z}$ is an auxiliary acceleration variable. With these in hand, the main iterations are detailed in {\bf{Algorithm}~\ref{algo1}}.
\vspace{-2mm}
\begin{algorithm}
  \caption{TF Domain LASSO Solver} \label{algo1}
  \begin{algorithmic}[1]
    \Require $\mathbf y_{\mathrm{TF}}$, $\mathbf{D}$, $\lambda$, $\varepsilon$
    \Ensure  Estimated TF domain fading vector $\hat{\boldsymbol{h}}$
    \Statex \hspace{-0.5mm}\hspace{-\algorithmicindent}\textbf{Initialization:} Initialize according to~(\ref{init_para})
      \For{$i=1,\dots$}
        \State $\mathbf{g}=\mathbf{D}^{\rm H}(\mathbf y_{\mathrm{TF}}-\mathbf{D}\mathbf{z})$
        \State $\hat{\boldsymbol{h}}^{i}=\psi\left(\mathbf{z}+\epsilon\mathbf{g},\gamma_{\mathrm{LASSO}}\right)$
        \State $\beta^{i+1}=\frac{1+\sqrt{1+4(\beta^i)^2}}{2}$
        \State $\mathbf{z}=\hat{\boldsymbol{h}}^{i}+\frac{\beta^i-1}{\beta^{i+1}}(\hat{\boldsymbol{h}}^{i}-\hat{\boldsymbol{h}}^{i-1})$
        \If{$\|\hat{\boldsymbol h}^{(i+1)}-\hat{\boldsymbol h}^{(i)}\|/
            \|\hat{\boldsymbol h}^{(i)}\| < \mathrm{\varepsilon}$}
        \State \textbf{break}
      \EndIf
      \EndFor
  \end{algorithmic}
\end{algorithm}
\vspace{-2mm}
\\ \noindent Note that the iterations terminate either when the maximum iteration count is reached or the relative change in the estimate falls below the tolerance $\varepsilon$. The operator $\psi(\mathbf{x},\gamma_{\mathrm{LASSO}})$ denotes the element‐wise complex soft‐thresholding function, whose expression is given by \vspace{-0.5mm}
\begin{align}
    \psi(\mathbf{x},\gamma_{\mathrm{LASSO}})=\max\left(0,1-\frac{\gamma_{\mathrm{LASSO}}}{|{x}_i|}\right)x_i, \forall x_i\in\mathbf{x}.
\end{align}
 Finally, we reconstruct the TF domain channel estimate $\hat{\mathbf{H}}_{\mathrm{TF}}$ by selecting each non-zero elements of $\hat{\boldsymbol{h}}$ and their corresponding delay and Doppler index estimates in $\hat{\boldsymbol{l}}$ and $\hat{\boldsymbol{k}}$, and then substitute them to (\ref{g_element})-(\ref{IO_TF}).

It is worth noting that, unlike the description on LASSO-based estimation in \cite{gaudio2021otfs}, which typically requires randomly placed pilots, our framework is not restricted to this assumption. This is due to the phase dictionary matrix $\boldsymbol{\Phi}(\boldsymbol{\tau},\boldsymbol{\nu})$ in (\ref{IO_TF_Xh}), which is constructed based on outputs of the DD-domain delay-Doppler estimates. Since the channel, including the delay and Doppler shifts, is inherently random, the resulting delay and Doppler estimates in the DD domain also exhibit randomness. As a result, the constructed phase dictionary matrix $\boldsymbol{\Phi}(\boldsymbol{\tau},\boldsymbol{\nu})$ naturally inherits this randomness. Therefore, there is no need to impose randomness on the pilot pattern in the TF domain, in contrast to the requirement in~\cite{gaudio2021otfs}.

\vspace{-1.2mm}
\section{Performance Analysis}
\vspace{-1mm}
\begin{figure}
    \centering
    \includegraphics[scale=0.27]{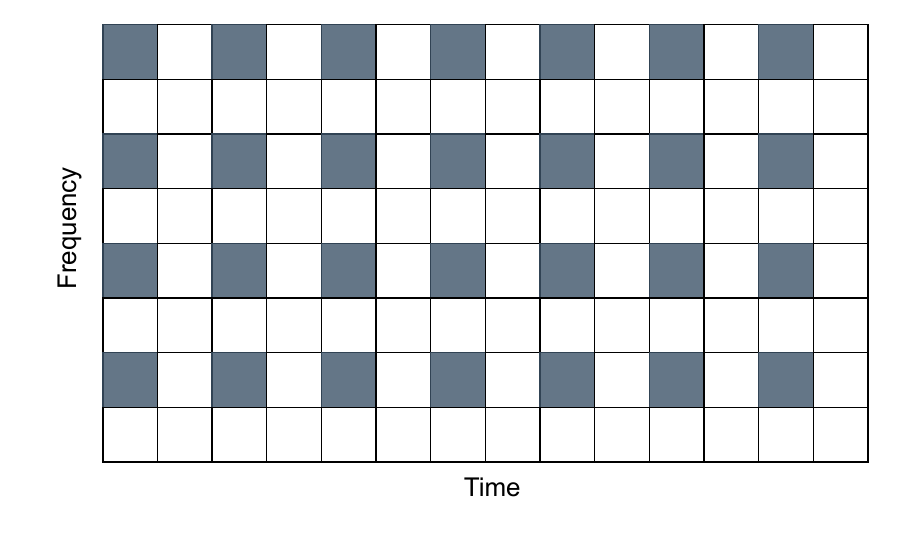}
    \caption{Pilot arrangement in the TF domain.}
    \label{fig:pilot}
       \vspace{-5mm}
\end{figure}

\begin{figure}[h]
	\centering  
	\subfigure[All-ones sequence.]{{\includegraphics[scale=0.22]{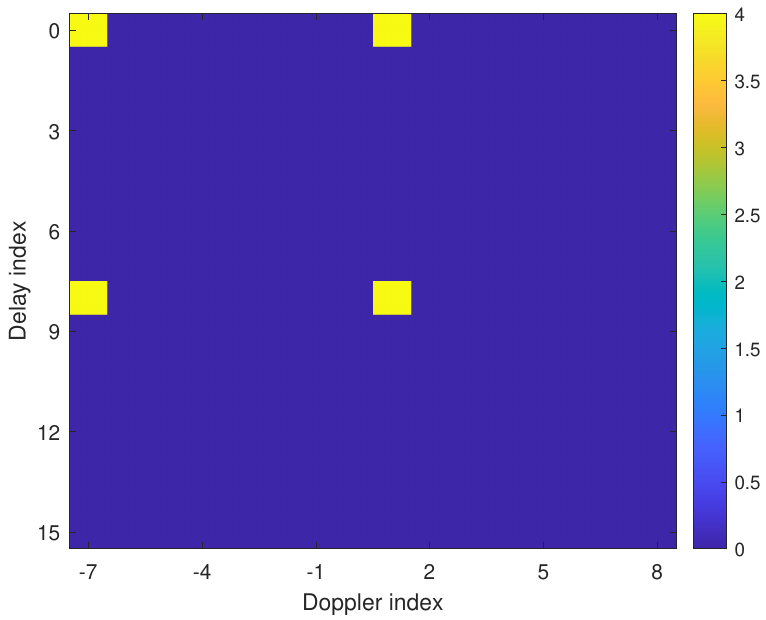}}}
    \subfigure[Walsh sequence.]{{\includegraphics[scale=0.22]{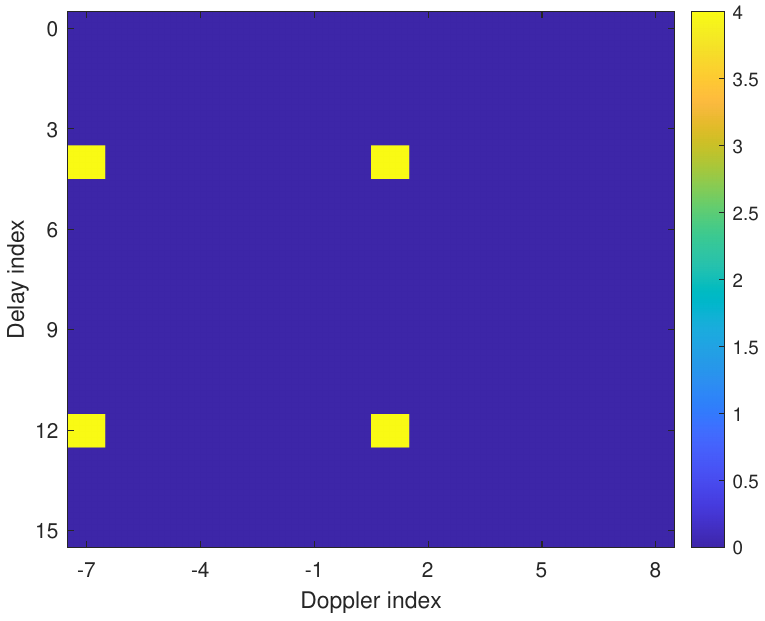}}}
    \subfigure[ZC sequence.]{{\includegraphics[scale=0.22]{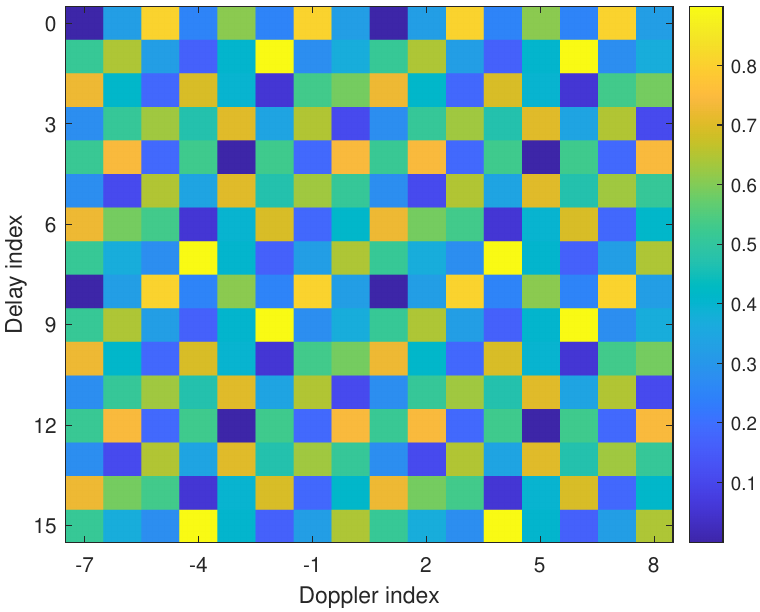}}}
    \caption{DD domain illustration of different TF domain pilot sequences ($M=N=16$).} 
        \label{fig:pilot_DD}
           \vspace{-4mm}
\end{figure}

\begin{figure}[h]
	\centering  
	\subfigure[All-ones or Walsh sequence.]{{\includegraphics[scale=0.335]{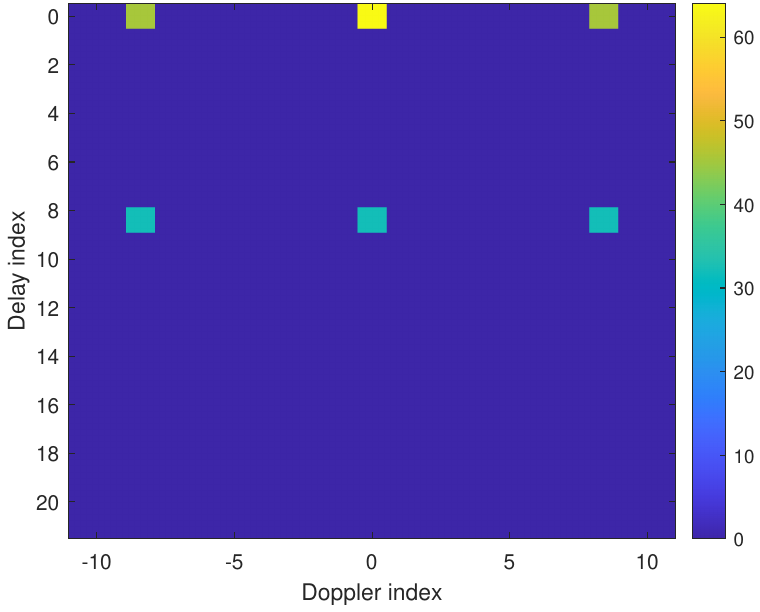}}}
    \subfigure[ZC sequence.]{{\includegraphics[scale=0.335]{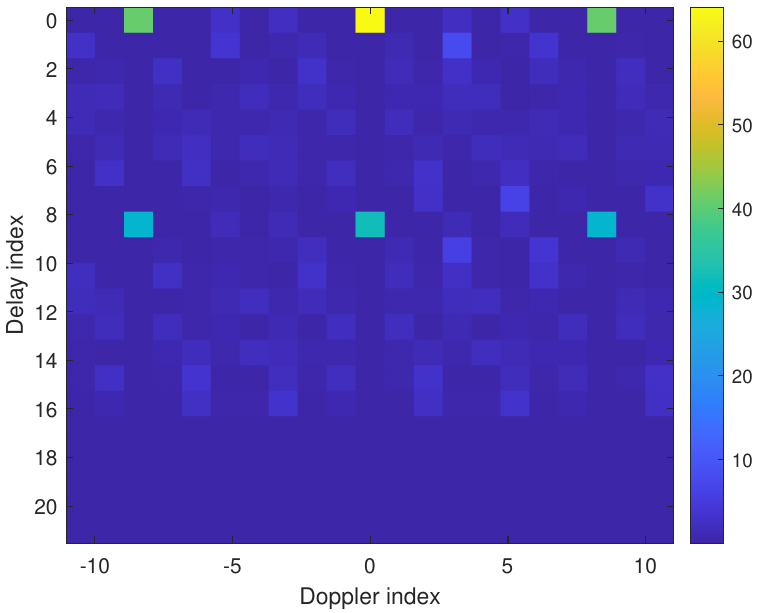}}}
    \caption{AFs of different TF domain pilot sequences ($M=N=16$).} 
        \label{fig:pilot_ACF}
           % \vspace{-8mm}
\end{figure}

In this section, we pursue a performance analysis for the proposed CDCE algorithm. It should be noted that the channel estimation performance of the proposed algorithm depends closely on the accuracy of the coarse estimation of the delay and Doppler shifts from the channel, which is also related to the underlying piloting scheme in the TF domain. Therefore, we focus on the performance evaluation of the coarse estimation of delay and Doppler in the following. Recalling~\eqref{DD_cross_corr}, let us define 
\begin{align}
    A_{\mathrm{DD}}[l,k]&\triangleq\sum_{l=0}^{M-1}\;
\sum_{k=0}^{N-1}
X_{\mathrm{DD}}^{*}[l',k']\,
X_{\mathrm{DD}}\!\bigl[[l'-l]_{M},\, [k'-k]_{N}\bigr] \nonumber\\
& \qquad\qquad \qquad \ \times \alpha[l'-l,k'-k]  e^{j2\pi\frac{k(l'-l)}{MN}},
\label{AF_twisted_conv}
\end{align}
where $\alpha[l'-l,k'-k]$ is given in \eqref{auto_corr}.  In fact,~\eqref{AF_twisted_conv} represents the discrete AF sample of ${\bf x}_{\rm DD}$ with respect to delay index $l$ and Doppler index $k$~\cite{li2024fundamentals}, and therefore the size of $\mathbf{A}_{\mathrm{DD}}$ is not necessarily bounded by $M$ and $N$.  
A pilot sequence with sharp AF characteristics facilitates more accurate peak
detection, which is particularly beneficial for estimating the delay
and Doppler shifts in our case. 
Therefore, in what follows, we focus specifically on this issue by examining the AF of the TF domain pilot sequence. 

We adopt a practical piloting scheme in the TF domain according to a lattice-type pattern\cite{cho2010mimo}, as shown in Fig.~\ref{fig:pilot}. Specifically, pilot symbols are periodically inserted along both the time and frequency dimensions to ensure reliable estimation in doubly selective channels. To shed light on the AF characteristics of different pilot sequences, we consider three representative types of pilot sequences: all-ones, Walsh, and Zadoff–Chu (ZC) sequences, whose equivalent DD domain representation is provided in Fig.~\ref{fig:pilot_DD}. 
From Fig.~\ref{fig:pilot_DD}, it can be observed that the all-ones and Walsh sequences remain well-localized after the domain transformation, while the ZC sequences spread across the entire equivalent DD domain grid. 

The discrete AF samples of the above three sequences are presented in Fig.~\ref{fig:pilot_ACF}, where the all-ones and Walsh sequences demonstrate a sharp peak with negligible sidelobes, which are beneficial for achieving high resolution in delay and Doppler estimation. In contrast, the ZC sequence exhibits a relatively high sidelobe value compared to that of the all-ones and Walsh sequences. Moreover, it is observed from Fig.~\ref{fig:pilot_ACF} that the all AFs exhibit certain periodic structure, which is determined by the underlying pilot pattern. For a practical design, the pilot sequence should be chosen such that the spacing between adjacent peaks exceeds the channel delay and Doppler spread, thereby avoiding ambiguity and aliasing in channel estimation. In addition, the characteristic of AF significantly influences estimation performance, particularly under data interference. A pilot sequence with poorly concentrated AF, i.e, weak main lobes and significant side lobes, would suffer from being drowned by data interference, thereby degrading the accuracy and robustness of channel estimation. \vspace{-2mm}

\section{Numerical Results}
\vspace{-1.5mm}

In this section, we evaluate the performance of our proposed CDCE design through numerical simulations, where both pilot-only case and practical case with coexisted pilot and data symbols are considered. Without loss of generality, we consider an OFDM system with $M=8$ and $N=14$, where the channel has $P=3$ independent resolvable paths with a maximum delay index $l_{\max}=2$ and a maximum Doppler index $k_{\max}=3$. For each path, the delay and Doppler indices are uniformly generated from $[0,l_{\max}]$ and $[-k_{\max},k_{\max}]$, respectively. The channel coefficients are independently drawn from a zero-mean complex Gaussian distribution with variance $\frac{1}{P}$. The signal-to-noise ratio (SNR) is defined as $\mathrm{SNR}=\frac{\mathcal{P}}{N_0}$, where $\mathcal{P}$ denotes the single pilot power. The average energy per transmitted data is also normalized to unity i.e., $E_s=1$.  The NMSE of channel estimate is defined as $\frac{||\hat{\mathbf{h}}_{\mathrm{TF}}-{\mathbf{h}}_{\mathrm{TF}}||^2_2}{||{\mathbf{h}}_{\mathrm{TF}}||_2^2}$, where $\hat{\mathbf{h}}_{\mathrm{TF}}=\mathrm{vec}\left(\hat{\mathbf{H}}_{\mathrm{TF}}\right)$ denotes the estimated TF channel vector and ${\mathbf{h}}_{\mathrm{TF}}=\mathrm{vec}\left({\mathbf{H}}_{\mathrm{TF}}\right)$ denotes the true channel vector.
According to our performance analysis, we consider the all-ones pilot sequence for numerical simulations. Additionally, for the DD domain thresholding, we set the threshold as $\gamma=\frac{\sqrt{N0}}{3}$ in the pilot-only scenario, and as $\gamma=\sqrt{\frac{\mathbf{v}^{\rm H}\mathbf{v}}{|\mathcal{R}|}}$ when data transmission is present, where $\mathbf{v}=\mathrm{vec}\left(\mathbf{V}_{\mathrm{DD}}\right)$ and $|\mathcal{R}|$ denotes the size of the target region. The parameters in LASSO solver are configured as $\lambda=0.01, \varepsilon=10^{-6}$, and maximum number of iterations $10^3$.

To evaluate the performance of the proposed CDCE method, we simulate several benchmark schemes. In particular, we consider single-tap least square (LS) and linear minimum mean square error (LMMSE), referred as ``ST-LS'' and ``ST-LMMSE'', respectively. These estimators are implemented by performing the following operations at the pilot positions:
\begin{align}
   \hat{\mathbf{h}}_{\mathrm{ST-LS}}&=\mathbf{y}_{\mathrm{TF}}\oslash\mathbf{x}_{\mathrm{TF}},\\
   \hat{\mathbf{h}}_{\mathrm{ST-LMMSE}}&=\frac{1}{1+\frac{1}{\mathrm{SNR}}}\odot\hat{\mathbf{h}}_{\mathrm{SW-LS}},
\end{align}
where $\oslash$ and $\odot$ denote the element-wise division and Hadamard product, respectively. The zero elements in $\hat{\mathbf{h}}_{\mathrm{ST-LS}}$ and $\hat{\mathbf{h}}_{\mathrm{ST-LMMSE}}$ are then recovered via linear interpolation based on the nearest non-zero elements in both the time and frequency dimensions.  Note that $\hat{\mathbf{h}}_{\mathrm{ST-LS}}$ and $\hat{\mathbf{h}}_{\mathrm{ST-LMMSE}}$ are of size $MN \times 1$, corresponding to the diagonal components of ${\mathbf{H}}_{\mathrm{TF}}$, where the ICI in ${\mathbf{H}}_{\mathrm{TF}}$ is not captured by those estimation schemes. For a fair comparison with CDCE, we reshape the estimated channel vector as $\hat{\mathbf{H}}_{\mathrm{ST-LS}}=\mathrm{diag}\left(\hat{\mathbf{h}}_{\mathrm{ST-LS}}\right)$ and $\hat{\mathbf{H}}_{\mathrm{ST-LMMSE}}=\mathrm{diag}\left(\hat{\mathbf{h}}_{\mathrm{ST-LMMSE}}\right)$. These reconstructed matrices are then compared with the true TF domain channel matrix $\mathbf{H}_{\mathrm{TF}}$ in terms of NMSE, after applying appropriate vectorization. Another conventional scheme that captures ICI is the full-sized LMMSE estimator, referred to as ``FS-LMMSE', which leverages \emph{a priori} statistical channel information and consequently, outperforms symbol-wise estimators. Its expression is given by
\vspace{1.5mm}
\begin{align}
    \mathbf{W}_{\mathrm{MMSE}}=\bar{\mathbf{C}}_{\mathrm{TF}}\mathbf{X}_{\mathrm{TF}}^{\rm H}\left( \mathbf{X}_{\mathrm{TF}}\bar{\mathbf{C}}_{\mathrm{TF}} \mathbf{X}_{\mathrm{TF}}^{\rm H}+N_0\mathbf{I}_{MN} \right)^{-1},\label{LMMSE} 
\end{align}
\vspace{-4mm}
\\ \noindent where $\mathbf{X}_{\mathrm{TF}}=\left(\mathbf{x}_{\mathrm{TF}}^{\rm T}\otimes\mathbf{I}_{MN}\right)$ denotes the equivalent pilot matrix corresponding to the vectorization of $\mathbf{H}_{\mathrm{TF}}$. Note that $\bar{\mathbf{C}}_{\mathrm{TF}}$ is the \emph{a priori} channel sample covariance matrix, which is obtained via Monte Carlo simulation. It is computed as $\bar{\mathbf{C}}_{\mathrm{TF}}=\frac{1}{K}\sum (\tilde{\mathbf{h}}_{\mathrm{TF}}-\bar{\mathbf{h}}_{\mathrm{TF}})(\tilde{\mathbf{h}}_{\mathrm{TF}}-\bar{\mathbf{h}}_{\mathrm{TF}})^{\rm H}$, where $K$ is the total number of Monte Carlo samples, $\tilde{\mathbf{h}}_{\mathrm{TF}}$ denotes the realizations of the TF domain channel, and $\bar{\mathbf{h}}_{\mathrm{TF}}$ denotes the \emph{a priori} sample mean of the TF domain channel, given by $\bar{\mathbf{h}}_{\mathrm{TF}}=\frac{1}{K}\sum \tilde{\mathbf{h}}_{\mathrm{TF}}$. Applying the estimator in (\ref{LMMSE}), we can obtain the \emph{a posteriori} estimate $\hat{\mathbf{h}}_{\mathrm{TF}}$ as~\cite{chong2025cross}   
\vspace{1.5mm}
\begin{align}
\hat{\mathbf{h}}_{\mathrm{FS-LMMSE,TF}} &\!= \!\bar{\mathbf{h}}_{\mathrm{TF}}+\mathbf{W}_{\mathrm{MMSE}} \left(\mathbf{y}_{\mathrm{TF}}-\mathbf{X}_{\mathrm{TF}}\bar{\mathbf{h}}_{\mathrm{TF}}\right),
% \hat{\mathbf{C}}_{\mathrm{d},\mathrm{TF}} &= \bar{\mathbf{C}}_{\mathrm{d},\mathrm{TF}}-\mathbf{W}_{\mathrm{MMSE}}\mathbf{X}_{\mathrm{d},\mathrm{TF}}\bar{\mathbf{C}}_{\mathrm{d},\mathrm{TF}}.\label{C_post}
\end{align}
\vspace{-4mm}
\\ \noindent It is important to note that the FS-LMMSE estimator involves a size of $MN\times MN$ matrix inversion, resulting in significantly higher computation complexity compared to symbol-wise approaches.  The last benchmark is TF domain LASSO estimator, referred as ``TF-LASSO'', which applies {\bf Algorithm~\ref{algo1}} without the \emph{a priori} knowledge from the DD domain. In this case, the length of $\hat{\boldsymbol{h}}$ in (\ref{IO_TF_Xh}) is set to the size of the DD domain frame $MN$, rather than the estimated number of paths $\hat{P}$ inferred from the DD domain. In other words, the estimator exploits all possible delay and Doppler shifts' combinations due to the lack of DD domain knowledge with increased complexity. Additionally, achieving good performance with this estimator typically requires a randomized pilot pattern, which may be impractical in real-world systems governed by current wireless standards on the pilot sequence.

\begin{figure}
    \centering
    \includegraphics[scale=0.35]{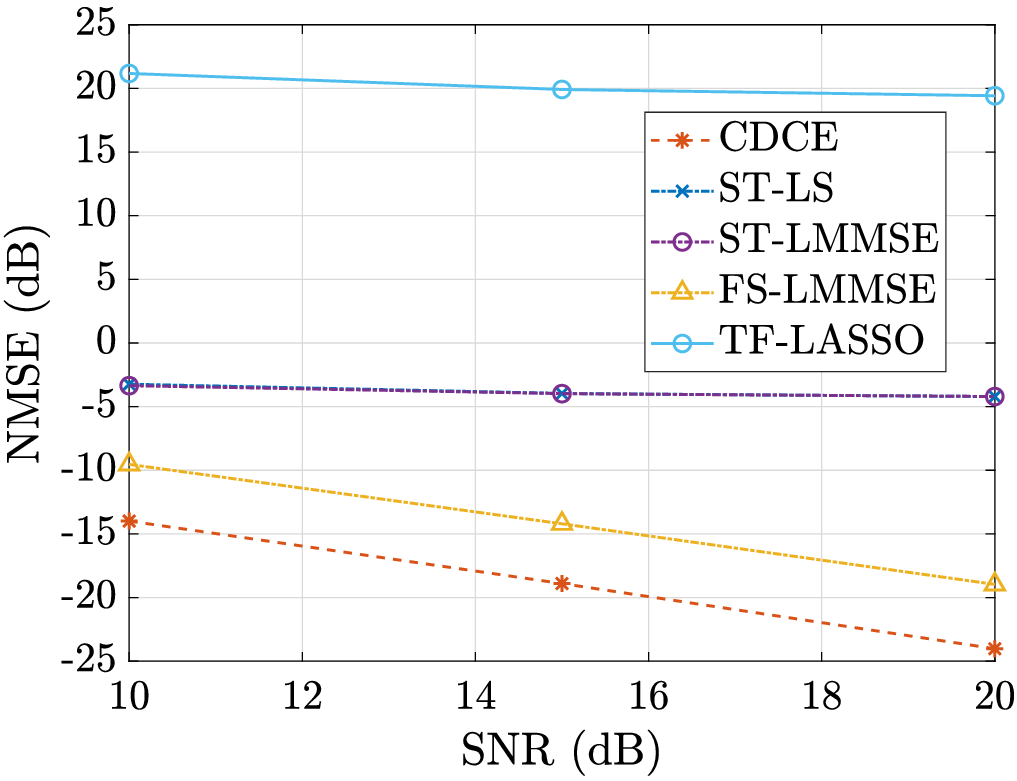}
    \caption{NMSE comparison for different estimation schemes.}
    \label{fig:NMSE}
       \vspace{-0.5cm}
\end{figure}

We present the numerical results for the pilot-only case in Fig.~\ref{fig:NMSE}.
It can be observed that the proposed CDCE achieves the best NMSE performance across various SNR conditions compared with all conventional TF domain schemes. Due to its reliance on a random lattice-type pilot pattern, the TF-LASSO scheme exhibits poor performance, with its NMSE remaining near $20$ dB. The ST-LS and ST-LMMSE estimators demonstrate a roughly flat performance around $-4$ dB across different SNR conditions. This arises from the fact that symbol-wise estimators are unable to capture the ICI induced by high-Doppler channels, resulting in a clear error floor. The FS-LMMSE estimator that mitigates those limitations achieves the best NMSE performance among all conventional schemes. Nevertheless, our proposed CDCE outperforms FS-LMMSE by approximately $4$ dB across all SNRs while requiring neither matrix inversion nor the \emph{a priori} channel informatio.

\begin{figure}
    \centering
    \includegraphics[scale=0.35]{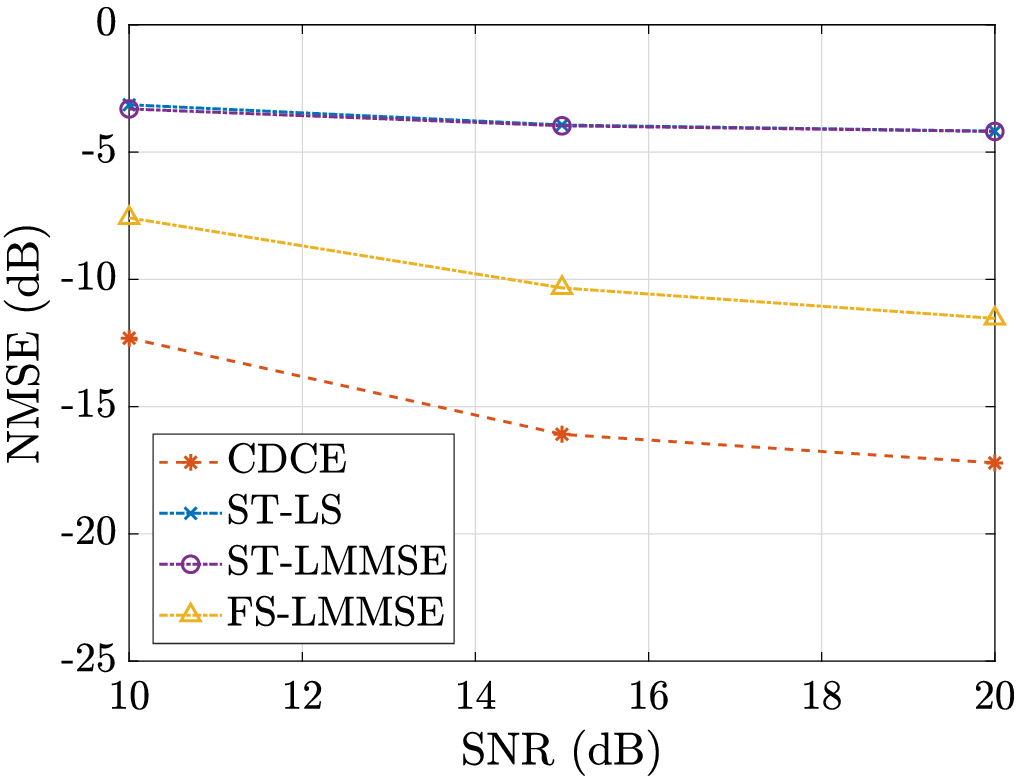}
    \caption{NMSE comparison for different estimation schemes with data.}
    \label{fig:NMSE_with_data}
    \vspace{-7mm}
\end{figure}

Fig.~\ref{fig:NMSE_with_data} illustrates an NMSE comparison under a realistic scenario in which pilots and data are transmitted simultaneously. The data symbols introduce substantial interference into the received pilot observations, magnifying the performance distance between the proposed CDCE and the conventional estimators. Specifically, the performance of ST-LS and ST-LMMSE again stays around $-5$ dB due to the dominant ICI and yields a persistent error floor. Although the NMSE of both FS-LMMSE and CDCE increases compared to the pilot-only scenario of Fig.~\ref{fig:NMSE}, CDCE now outperforms FS-LMMSE by approximately $5$ dB across the full SNR range. This robustness is due to that the DD domain correlation operation is able to attenuate a significant portion of data-induced interference before returning the estimate back to the TF domain.  Furthermore, it is worth highlighting that the conventional TF-LASSO scheme lacks the capability to mitigate data interference during channel estimation. In particular,  the LASSO solver attempts to fit the estimate $\hat{\boldsymbol{h}}$ to the received signal that contains both pilot and data components, rather than the pilot-only receive signal. This mismatch leads to a biased estimate, causing $\hat{\boldsymbol{h}}$ to deviate substantially from the true channel.

\vspace{-1mm}
\section{Conclusion}
\vspace{-1mm}
In this paper, we proposed a novel cross-domain channel estimation scheme for conventional OFDM systems by availing of the advantages of both the DD domain and the TF domain. The proposed approach first performs DD domain correlation to extract estimates of the delay and Doppler shifts, which are then passed to the TF domain, where the fading coefficients are estimated via an LASSO solver. Thanks to the interference-resilient nature of DD-domain correlation, the scheme can reliably estimate the delay and Doppler parameters even in the presence of data interference. Moreover, by leveraging the \emph{a priori} delay and Doppler shift estimates obtained from the DD domain, the TF domain LASSO solver obtains a low computation complexity, where the estimation process is limited to only the identified delay and Doppler shift combinations, rather than an exhaustive search over all possible combinations. Simulation results show that the proposed scheme consistently outperforms conventional OFDM estimators in terms of NMSE, achieving at least a $4$ dB improvement in pilot-only transmission and a $5$ dB improvement in data-pilot transmission, without requiring matrix inversion or any \emph{a priori} channel information. 

% \bstctlcite{IEEEexample:BSTcontrol}
\vspace{-2mm}

\bibliographystyle{IEEEtran}
\bibliography{ref}
\vspace{-8mm}

% \newpage

% \vfill

\end{document}